# An Overview of the AAVSO's Information Technology Infrastructure From 1967 to 1997

**Richard C. S. Kinne**
*AAVSO Headquarters, 49 Bay State Road, Cambridge, MA 02138; rkinne@aavso.org*

**Abstract** Computer technology and data processing swept both society and the sciences like a wave in the latter half of the 20th century. We trace the AAVSO's usage of computational and data processing technology from its beginnings in 1967, through 1997. We focus on equipment, people, and the purpose such computational power was put to, and compare and contrast the organization's use of hardware and software with that of the wider industry.

## 1. Introduction

There are some organizations and purposes for which data processing and computers seem to have been tailor-made. One of these is the AAVSO. To their credit, the organization's leaders, specifically Directors Margaret Mayall and, later, Janet Mattei, realized this synergy fairly early on. During the AAVSO's Annual meeting in 1966, Mayall consulted with AAVSO member Professor Owen Gingerich at the Smithsonian Astrophysical Observatory (SAO) about the feasibility of digitizing the AAVSO's variable star observational data (Welther 1970). In 1967, with the cooperation of SAO, Mayall and the AAVSO began digitizing observations for what would become the AAVSO International Database. This was followed, in fairly short order, by the start of custom programming, again with the cooperation of SAO, to analyze the observational data being compiled (Welther 1970).

Along with other organizations of the AAVSO's size and non-profit mission, the 1980s were a transitional decade that saw computer technology used not just for data, programming, and analysis (although that certainly continued), but also for routine office work. The 1990s, for both the AAVSO and the wider world, was the "decade of networking," with Janet Mattei overseeing both the installation and use of a local area network, and the marriage of the organization to the then new World Wide Web.

## 2. Early developments

As early as 1958, Margaret Mayall considered "the possibility of processing all variable star data at headquarters by means of IBM punched cards" (Williams and Saladyga 2011); but it was not until 1966 that Mayall and the AAVSO Council began to bring that idea into fruition. During the AAVSO's Annual Meeting in 1966 Gingerich, Mayall, and the AAVSO Council discussed how this might occur. Gingerich, in what would become the first of a long line of important contributions to the AAVSO, believed he could donate some free access time on the SAO computers to the AAVSO. All that the AAVSO would need to invest in, he said, would be a card punch (keypunch) machine. At the May 1967 meeting, the AAVSO Council decided to rent a keypunch machine to begin a critical process of remaining relevant and moving to increased professional acceptance of the organization (Saladyga and Williams 2011).

Prior to this decision, observations were recorded in ledger books, and light curves were hand-plotted from these data for particular, special, or requested stars. Incoming data started to grow in leaps and bounds at this point and new ways had to be found to graphically represent the data. Data digitization enabled plotting and analysis programs to be used, which greatly enhanced the usefulness of the data for both amateur observers and professional astronomers alike. Once the AAVSO crossed this Rubicon, going back was unthinkable.

Digitized cards had other uses that observations in ledgers could never approach. Cards could be handled in discrete batches, sorted automatically based on criteria, and could serve as input to computer programs.

Once photometric data were digitized, it was easy to find reasons to use them in that form. Over the next few years, for example, the AAVSO made an effort to use computer programs and software plotters to help automate and produce compilations of observations and light curves of selected stars—the *AAVSO Report* series—using the keypunched data that had been collected.

By 1971 the AAVSO had digitized one million observations. There were millions more, but it was decided at that time that digitizing current observations as they came in would take precedence, assuming that astronomers would need current data more than archival data.

By 1973 keypunch machines were common in large office and scholastic environments. Charles Scovil, for example, made arrangements with Darien High School in Connecticut to use their keypunch machines during off hours (Scovil 1972). The AAVSO developed its own keypunch training program which enabled volunteers and staff to enter the observations that came in each month (Figure 1). Some observers were already starting to send observations to AAVSO Headquarters on keypunched cards (Mayall 1973).

In just six years the number of cards being processed and physically stored by the AAVSO—then at about 100,000—was becoming a problem (Mattei 1974). At this point only data from 1960 on, with the exception of a two-month period in 1973, had been digitized; the project of digitizing AAVSO observations from 1911 to 1960 had not yet begun. *Report 30* was being compiled, and the 130,000 published observations used for *Report 28* and *Report 30* were stored on four separate magnetic tapes kept in four separate locations for safety. With those tapes the AAVSO had a small taste of the future, and was gearing up for its second data processing revolution.

The AAVSO had digitized its membership list by 1973 as well, allowing the office to create computerized mailing labels and to generate selective mailings (Ford 1974; Mattei 1974). Data analysis programs had been part of the AAVSO from the start of its digitization process created at that point by AAVSO member Barbara Welther, who was a staff researcher at SAO. One program written by Welther during this period was one that found and noted maximum and minimum dates for Mira variables for *Report 30*. Up to this point the AAVSO staff were processing these data by hand, augmented by a program that produced 10-day mean light curves. It now became faster to do it all by computer, and Barbara Welther came up with the program with which to do it.

**3. The second revolution—better control of data**

1974 saw its first of many AAVSO in-house programmers when MIT student Richard Strazdas joined the staff (Figure 2). Under Strazdas the AAVSO continued to move away from the hand-plotting of light curves, a road which eventually led to the web-based Light Curve Generator that we know today. Strazdas' method involved deriving light curves from density curves where the number of observations at specific magnitudes were printed at each date. This method, for the first time, allowed the AAVSO staff to easily find observational outliers, notifying and guiding observers toward gathering better data. At this point light curves were produced using alphanumeric characters on line printers. Their resolution was quite poor. It was not possible to plot individual observations, only 5-day means, using this technology (Mattei 1975).

Over the next four years Strazdas wrote several programs in FORTRAN (a language the AAVSO still uses productively today) that specifically used data that were stored and read from magnetic tapes. The data processing procedure began with observations being keypunched onto cards, which were then stored on magnetic tape. A program called VALID initially checked the data and corrected or flagged it for errors in designation, star name, and so on. BSORT then read the output of this program, which was also stored on tape, taking the place of a mechanical card sorter. A third program, BMERGE, combined the two different sorted data sets. Thus the first half of the second data revolution for the AAVSO had been accomplished. Instead of using cards for computer program input, the cards were now a backup to the much more flexible magnetic tapes (Hill 1977).

The second half of the second revolution involved two computer plotting units then owned by SAO; a Versatec electrostatic plotter, and a Calcomp ink-based plotter. These plotters, under the direction of Strazdas-written programs, aided by Robert S. Hill, allowed for the first time individual data points to be plotted as the computerized light curves that we would recognize today. With such improved resolution in computer produced light curves, the observational density and scatter in a plot of observations (always evident on the data hand-plots) finally became apparent (Ford 1977). This method was used for all future reports and publications, thus completing the AAVSO's second data processing revolution (Hill 1977). It had started with punch card processing and gone to tape processing, and from line-printer produced 10-day mean light curves to plotted individual observations. The plotting aspect of the revolution, while seemingly starting out well, had a hard birthing.

Technology continued to move forward in late 1978, but the AAVSO had to halt for a technological pit stop. The Harvard-Smithsonian Center for Astrophysics (CfA) upgraded its main computer from a CDC 6400 to a DEC VAX 11/780 and all the programs that ran on the CDC 6400 had to be rewritten for the DEC VAX architecture. Having no full-time programmer at the AAVSO, this conversion took weeks—it was supervised by Richard Strazdas with the help of two students, Christopher Walton and Sandra Galejs. The switchover put the publication of *Reports 38* and *39* on hiatus while Strazdas and his team converted the needed programs and developed new ones (Mattei 1979).

Data entry, something the AAVSO had gotten rather good at, forged ahead through mid-1980. Under Elizabeth Waagen's direction, all data from 1960 up to the then current time—325 boxes of IBM punch cards comprising 650K of data—were now on magnetic tape and sorted by star name and date. Light curve plotting stumbled, however, and the *Reports* could not be published. The Calcomp plotter that Strazdas had written his plotting programs for had never been moved to the new DEC VAX 11/780 from the CDC 6400. The AAVSO purchased a new plotter—an FRS80 Graphics computer from AVCO Computer Services—and Strazdas adapted his plotting programs to it (Mattei 1980).

By mid-1981, with the data from 1960 onward now machine-readable, progress towards the goal of converting into machine-readable format all data from the founding of the AAVSO to 1960—2.5 million observations—began (Mattei 1981). With this project, and the need to continue keeping up with incoming observations, the AAVSO was pushed into its next revolution, its largest yet: independence.

**4. The first in-house computer system**

The technology and cost of microcomputers had just gotten to a point where they might be a feasible alternative for the AAVSO. At the other end of the scale, the increasing volume of IBM punch cards was literally filling Headquarters and squeezing everything else out. An in-house system was needed that could deal directly with floppy disks and maintain the publishing schedule the AAVSO had created. Mattei initiated a massive research and funding project to find and purchase an appropriate and affordable microcomputer system. It culminated in the AAVSO obtaining, through a grant from the Research Corporation, two

Ithaca Intersystems computers in December of 1981 (Mattei 1982). The first was a Z80-based computer running CP/M, with a graphics terminal and a plotter. The other was the DPS-8000, a Z8002-based multi-user system running COHERENT, a UNIX look-a-like operating system, with three terminals for data entry, word processing, and other office work (Figure 3). Both systems boasted 64,000 bytes of random-access memory.

The acquisition of its own computer did not immediately cut the AAVSO's ties to Harvard—not by a long shot. While the Ithaca Intersystems computers were advanced microcomputers for the time, they were too small to handle the AAVSO's data processing needs. The Ithaca's greatest contribution was that it enabled the AAVSO to move past punch card storage to eight-inch floppy disks for temporary data storage. Now, instead of data being punched onto cards which were stored and then retreived to be read onto magnetic tape, data were keyed onto the disks, then verified (re-keyed to check for errors), then converted to a DEC-readable format, read into the PDP 11/60, transferred to the DEC VAX 11/780 for processing, and stored on permanent tape, while storing the diskettes as a backup.

The monthly data inflow to the AAVSO—15,000 to 20,000 observations at this point (Waagen 1984)—was too much for the microcomputer to handle; observations were still stored on magnetic tape which the AAVSO could not read on its own. When observations of a specific time period were needed for publication, the storage tape would be read into the VAX, transferred to the PDP, and copied onto diskettes for processing at the AAVSO. In 1984 the PDP 11/60 was decommissioned and its disk readers transferred to the VAX 11/780, taking one step out of this process (Waagen 1984).

While diving into computer use itself, the AAVSO also recognized that its observers were able to take advantage of this technology as well. To assist them, the AAVSO sponsored a computer workshop as part of its AAVSO 73rd Spring Meeting in 1984, in Ames, Iowa.

Despite the advances in information processing, the huge *Reports* were abandoned as Mattei learned that researchers preferred a long span of data on one star to a short span on hundreds. Capitalizing on the information technology that it did have, the AAVSO began publishing a *Monograph* series, each of which concentrated on the twenty-year light curve of a specific star. The International Astronomical Union (IAU) welcomed and praised this initiative (Mattei 1984).

**5. Growth in data processing capability and application**

In 1986 the AAVSO moved to Birch Street and prepared to celebrate its 75th Anniversary. As one can imagine, the move put most work on hold for awhile as things were packed, moved, and unpacked (Mattei 1986). Still, the staff, under Mattei's leadership, continued to gain technological ground. A Perkin Fund grant enabled the hiring of two full-time staff for the archival data entry project. With their help, by 1986, twenty-five percent of the AAVSO archival data for 1911–1961 had been converted to machine-readable form.

The IBM PC clone, and the first stages of networking, came to the AAVSO in 1987. The clone, sporting a 40-megabyte hard drive, connected the AAVSO to CfA through a modem device. The Kenilworth Fund bought Headquarters a laser printer and scanner. Observers began submitting data to Headquarters using diskettes and email. By 1989 the first articles featuring computer analysis of variable star data by AAVSO members were being published in *JAAVSO* (Mattei 1988). FORTRAN programs originally written for the VAX 11/750 were now rewritten for the IBM PC. Also, Grant Foster (Figure 4) began to write a series of graphical programs which allowed real-time manipulation of light curve data on the computer screen; these programs were not for data entry and editing, but for actual statistical analysis of the data (Mattei 1989).

The addition of 600 megabytes of hard drive space on the main computer in 1990 allowed all the variable star photometry from 1960 onward to come home from CfA. AAVSO staff migrated the data from storage tape to magnetic cartridges. The AAVSO installed its first local area network (LAN) in 1991 using 10base-2 LANtastic technology. These were used to tie together ten PC clones bought for the staff through a NASA grant (Mattei 1991). Headquarters began experimenting with commercial data services by putting astronomical data on Compuserve in 1992 (Mattei 1992).

A Theodore H. Dunham Fund for Astrophysical Research grant expanded the hard drive storage capability at AAVSO Headquarters to 2.4 gigabytes in 1993, just in time to aid in the completion of the archival project. Now the AAVSO had the entire AAVSO International Database in computer readable form right on site! Spearheaded by Grant Foster, AAVSO staff wrote programs to facilitate analysis of the data that Headquarters had spent more than twenty years digitally archiving.

In 1995 William Mackiewicz (Figure 5) became the AAVSO's first webmaster; he created the organization's first website and file transfer protocol (FTP) server. An IBM PC clone running GNU/Linux provided the AAVSO's first Internet services. By 1997, the AAVSO used its website to provide charts, and its *AAVSO News Flash*, *Circulars*, and *Alert Notices* to the public. With over 400 visits a day, the AAVSO website was named one of the top education-related sites on the Net (Mattei 1998). Users responded in kind with fully fifty percent of the monthly reports being sent to Headquarters electronically by 1997.

The AAVSO went from one computing strength to the next, but there were a few potholes along the way. Increasing reliance on technology meant that problems would crop up from time to time, and the AAVSO was not immune to this. In 1991 a bad sector on a hard drive caused the first AAVSO data loss. Through redundant diskette backups the staff was ultimately able to recover the data. In 1997 a vandal broke into the Linux server but did not compromise data. The vandal only created and ran his own chat room.

**6. Successfully riding the technological wave?—an assessment**

It seems clear that the AAVSO had a good track record of using technology to accomplish its mission. How close was the AAVSO to "riding the technological wave" that confronted it? Some non-profits don't do well with this, usually due to limited funds.

It is difficult to assess the exact state of a technological wave in a practical sense. Keypunch machines and keypunch cards were in use before WWII. The AAVSO started using them in the mid-1960s, borrowing time and resources from larger organizations to build its computational legacy. The AAVSO's computational technology from the mid-1960s until 1980 was dependent on the resources used at SAO and CfA, so during this time how the AAVSO fared technologically was somewhat tied to how those organizations fared.

The AAVSO moved to punch cards at the very end of their practical life. For programming purposes punch cards had fallen out of use in production environments by the 1970s, but they would continue to be used for data storage at the AAVSO right through the early 1980s, largely due to the availability of older machines in large data centers. In the end, the AAVSO was driven from cards for the exact same reason everyone else was—lack of space.

The AAVSO, through its partnership with CfA, kept up with hardware advances pretty well with the VAX 11/780 mini-computer. CfA upgraded to this computer in 1978, less than a year after DEC announced it at the Annual Shareholder's Meeting in 1977 (Digital Equipment Corp. 1997).

Sometimes being close to the edge can have its downside if looked at in hindsight. The AAVSO spent a good bit of time and research toward purchasing their two Ithaca Intersystems computers. To modern eyes the purchase of a CP/M system in 1981 looks shortsighted, but at that point there really wasn't any other microcomputer option available. The very first IBM PC went on sale in August of that year and had no track record as yet. Furthermore, DOS was not designed as a multi-user operating system, or as a file server. CP/M had over a ten-year history and, indeed, IBM itself had originally selected CP/M as the operating system of the IBM PC, but talks in 1980 with Digital Research, Inc. failed, and IBM decided to go with with Microsoft for its operating system (Anthony 2011).

While perhaps the best choice, the Ithaca Intersystems computers also featured a swan song in terms of storage. The system initially used eight-inch floppy disks introduced for CP/M in 1977. This was the last introduction of an eight-inch floppy drive. While old technology, the drive featured one megabyte of storage formatted for CP/M, while the best a 5.25-inch floppy could do at the time was 87.5 kilobytes (Sollman 1978).

The first IBM-compatible was released in late 1982. Several companies struggled for a year or so before finally achieving an acceptable level of compatibility with it (Reimer 2005). It took until 1987 for PC-compatible computers to show up at AAVSO Headquarters, by which time they had become commodities. In this case the AAVSO waited five years to enter the PC market. In parallel, that first AAVSO PC-compatible allowed communication with the CfA through a Hayes Smartmodem compatible, which was released in July of 1981 (Markoff 1983).

In contrast, in terms of local area networking, once the PCs arrived at Headquarters, the AAVSO stepped right into setting up a LAN. While Artisoft's LANtastic is not widely remembered today, at that time it rivaled Novell in the PC networking market. Neither Novell nor Artisoft foresaw the rise of TCP/IP networking, but both products still exist today. LANtastic is currently on version 8.

Arguably, one of the most significant information technology events for the AAVSO was its adoption of the World Wide Web. Sir Tim Berners-Lee released the Web in August of 1991 (Berners-Lee 1991). The AAVSO's first web server went online in 1995. While a four-year lag may seem somewhat significant, Berners-Lee's Web did not take off until the introduction of the Mosaic web browser in 1993 (Andreessen 1993).

In the same year that Berners-Lee introduced the Web, Linus Torvalds introduced the Linux kernel (Torvalds 1991) which the AAVSO's first, and all subsequent, web servers ran on. Torvalds' release of the kernel under the GNU General Public License in 1992 accelerated the creation of the free UNIX-like operating system which we know today as GNU/Linux (Stallman 1997). In March 1994 Linux reached version 1.0 and Linux distributions such as Slackware and Debian were in wide release. The AAVSO adopted GNU/Linux just over a year after it became practical.

**7. Conclusion**

While the AAVSO is a non-profit corporation, it may not be valid to compare them to other non-profits such as libraries in their technological adoption curve. At its heart the AAVSO is a technological organization and so it needs to come up to a higher bar. Couple this with the financial issues that most non-profits seem to go through—and the AAVSO is no stranger to financial challenges—the organization seemed to do a pretty impressive job of taking advantage of technology whenever it could.

Taking advantage of technology when it became available requires adaption to change at a very fundamental level. Both people and organizations find that difficult. It takes strength in an individual and strong leadership in an organization. Margaret Mayall, with the help of technologically astute people on the AAVSO Council at the time such as Clint Ford, as well as friends at SAO such as Owen Gingerich and Barbara Welther, allowed the AAVSO to make its initial leaps into using technology to improve the efficiency of the organization.

When Janet Mattei initially came on board she continued with the progress that Mayall had begun. Soon, though, spurred on by the success of the initial digitization project that allowed her to reach for larger government contracts and backing, Mattei made significant steps of her own that not only continued to improve the efficiency of the organization, but allowed it be stay competitive and relevant in the face of progress.

**8. Acknowledgements**

The author would like to acknowledge Elizabeth O. Waagen, AAVSO Senior Technical Assistant, and Michael Saladyga, AAVSO Technical Assistant, for serving as co-authors on the poster that preceded this paper. Dr. Saladyga's assistance with the AAVSO Archives was invaluable. We would like to acknowledge the rest of the AAVSO staff as well, especially Dr. Aaron Price, for their inspiration and support.

Table 1. Timeline of events in the development of AAVSO information technology.

| Date | Event |
|---|---|
| 1967 | Computer processing starts for the AAVSO using facilities at the Smithsonian Astrophysical Observatory to put data on IBM punch cards. |
| 1972 | Charles Scovil makes arrangement with Darien (Connecticut) High School to use its key punch machine in off hours. With that help, the AAVSO staff is working on keypunching incoming observations and working on starting work on reports from 1911 and later. |
| 1973 | The AAVSO membership list information is now put on IBM punch cards. The main data processing thrust at this point is using keypunched entered data in preparing the *Reports*. At this point *Report 30* is being compiled. The published data for *Reports 28* and *29*—130,000 observations—are being put onto four copies of magnetic tape. |
| 1975 | MIT student Richard Strazdas develops (based on an existing program) a method wherein light curves are obtained as density curves in which the number of observations at specific magnitudes are printed at each date. The program then plots the light curve. This allows the study of computerized plots and the detection of anomalous observations. Observational data from 1960 to May 1968 are processed. June 1968–November 1974 is not processed or must be reprocessed due to error. This became known as "The Gap." AAVSO staff computerizes the membership database and mailing labels for mailings which are made using SAO computers and printers. |
| 1978 | Harvard-Smithsonian Center for Astrophysics (CfA) upgrades its CDC 6400 computer to VAX 11/780. The AAVSO converts all data and programs to be compatible with the DEC VAX 11/780. The PDP 11/60 is still in use there as a data reader. |
| 1979 | All data from 1960, sorted by star and date, are now on magnetic tape and are machine-readable. |
| 1980 | The "Gap Data" are finally processed. The AAVSO begins computerization of data from 1911 to 1961. This is a multi-year project. AAVSO Headquarters is taken over by punch cards; Director Mattei is determined that something needs to be done about their storage. The AAVSO researches the feasibility of purchasing its own computer system using 8-inch floppy diskettes as storage media. An in-house system is needed to offset increasing publishing costs. The |

Table 1. Timeline of events in the development of AAVSO information technology, cont.

| Date | Event |
|---|---|
| | system needs to have a graphics terminal, plotter, and printers and be compatible with the DEC VAX at CfA. |
| 1981 | Through the Charles M. Townes Fund, the AAVSO buys two Ithaca Intersystems microcomputers with the CP/M operating system. One is a single-user system comprised of a computer, terminal, graphics terminal, and plotter which is used to plot data on screen, check, edit, and plot the data to paper. The other is a multiuser system with three terminals, two for data entry, and one for word processing for *JAAVSO*, correspondence, mailing list, and other office work. Incoming observations are now stored on 8-inch disks and processed using the VAX at CfA, and stored on magnetic tape at CfA.. |
| 1982 | AAVSO Treasurer Theodore Wales buys a terminal and a pair of disk drives for the new AAVSO computer system. The monthly inflow of observations attains the 15,000–20,000 level—too big for the Intersystem computer to handle. These data still processed at CfA. |
| 1984 | CfA decommissions its PDP 11/60. The disk readers are put on the VAX allowing the VAX to read AAVSO data directly. Charles Jones, an MIT student, writes a data editing program for the Intersystems computer allowing editing to be done in-house. The AAVSO holds a Computer Workshop as part of its 73rd Spring Meeting. |
| 1985 | 25% of archival data from 1911 to 1960 is put to tape. HQ uses its computers to produce the *AAVSO Monograph* series. |
| 1986 | The AAVSO moves to Birch Street. HQ begins exploring the possibility of observers submitting data on diskettes or via modem. There is a near-complete turnover in AAVSO programming staff. |
| 1987 | A new IBM PC connects AAVSO HQ with the DEC VAX at CfA via modem. The PC has a 40 megabyte hard drive. The Kenilworth Fund buys HQ a laser printer and scanner for the PC clone. |
| 1989 | The first *JAAVSO* articles detailing computer use in amateur variable star observation and research begin appearing. VAX FORTRAN programs are rewritten to run on PC clone. Data processing is now done at HQ, not CfA, but CfA equipment is still used for tape storage. The AAVSO begins supporting the HIPPARCOS data mission. The AAVSO researches data storage solutions with the goal of migrating |

Table 1. Timeline of events in the development of AAVSO information technology, cont.

| Date | Event |
|---|---|
|  | all CfA-stored data to in-house storage. The archival data project is 77% complete. Grant Foster writes a new light curve plotting program that uses a scale compatible with existing hand-plotted light curves. |
| 1991 | NASA grants provide a terminal or stand-alone computer system (IBM clone 186, 386, 486) for each staff member (ten in all). All workstations are networked via LANtastic LAN to the main computer for file access. First reported data problem: bad sectors on a disk cause data loss that needs to be recovered. The archival data project is 97% complete. |
| 1992 | Grant Foster writes programs to plot light curves on-screen for any star, expand any portion of the light curve, identify observations of observers on the light curve, and evaluate an observation and change its status. The AAVSO is now listing data on Compuserve. |
| 1993 | AAVSO staff complete the data entry phase of the archival data project. Now the data have to be processed! The plan is to have this done in three years. A Dunham Grant adds 1.8 gigabytes of storage to the main computer system bringing its total to 2.4 gigabytes. The AAVSO now switches its focus somewhat to writing programs to analyze its data. |
| 1994 | The AAVSO purchases its first Pentium computer and CD-ROM reader through a NASA HIPPARCOS grant. |
| 1995 | The AAVSO appears on the World Wide Web. William Mackiewicz, the AAVSO's first webmaster, also establishes an FTP site. Internet services are being run on a PC clone using GNU/Linux. |
| 1996 | The AAVSO acquires two Pentium computers, and places 114 charts on its FTP site. The AAVSO website sees about 228 visits per day. |
| 1997 | The AAVSO uses its website to distribute the *AAVSO News Flash, Circular*, and *Alert Notices*. The website now sees 483 visits per day. The FTP site has 2,179 files downloaded each month. The entire AAVSO database is archived on ZIP disks. 50% of monthly observing reports arrive electronically, up from 32% the previous year. Archival processing completed. Grant Foster writes WWZ, a time-series analysis program. All workstations are running Windows95 and are upgraded to 486s or Pentium. A vandal breaks into the GNU/Linux server. The AAVSO website named one of the best education-related sites on the web. |

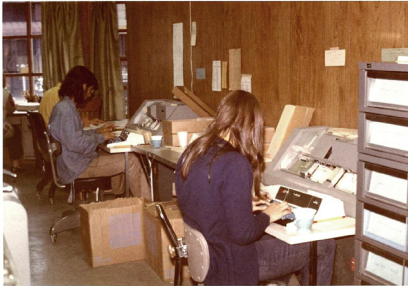

Figure 1. Keypunching operations performed by work-study students at AAVSO's Concord Street Headquarters in the early 1980s.

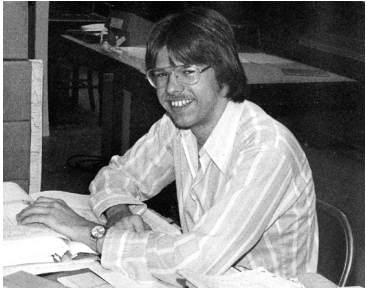

Figure 2. Richard Strazdas, MIT student who wrote data processing and file-transfer programs for the AAVSO beginning in 1974.

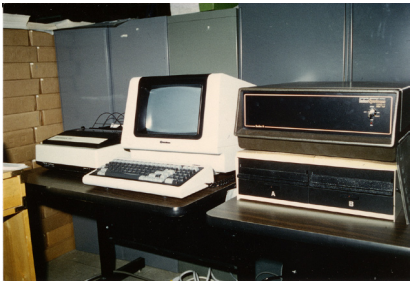

Figure 3. The Ithaca-Intersystems computer at AAVSO's Concord Street Headquarters, early 1980s. The system brought AAVSO's data processing operations in-house. Some of the hundreds of boxes of punch cards can be seen on the left, forming a work-area partition.

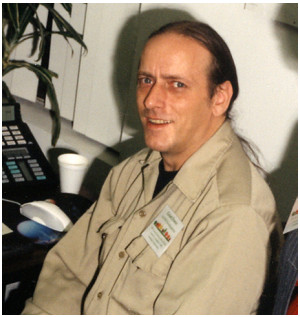

Figure 4. Grant Foster, AAVSO programmer from the late 1980s to the early 2000s.

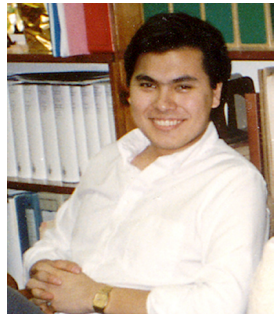

Figure 5. William Mackiewicz became the AAVSO's first webmaster in 1995.